\documentclass[amsmath,amsfonts,aps,twocolumn,twoside,superscriptaddress,longbibliography]{revtex4-1}

\usepackage{palatino} 
\usepackage{mathpazo}
\usepackage{graphicx}

\usepackage[utf8]{inputenc}
\usepackage[bookmarks]{hyperref}
\hypersetup{colorlinks=true,citecolor=blue,linkcolor=blue,filecolor=blue,urlcolor=blue}
\usepackage{amsmath,amssymb,amsthm}
\allowdisplaybreaks[4]
\usepackage{dsfont}

\usepackage{cleveref}

\usepackage{autonum}

\usepackage{enumerate}

\usepackage{mathtools}

\usepackage{tikz}
\usepackage{pgfplots}
\usepackage[caption=false]{subfig}
\usepackage{xcolor}
\definecolor{orange}{HTML}{CC7A00}
\usepgfplotslibrary{fillbetween}

\definecolor{dgreen}{HTML}{00CC00}

\newcommand{\cA}{\mathcal{A}}
\newcommand{\cB}{\mathcal{B}}
\newcommand{\cC}{\mathcal{C}}
\newcommand{\cD}{\mathcal{D}}
\newcommand{\cE}{\mathcal{E}}
\newcommand{\cF}{\mathcal{F}}

\newcommand{\cH}{\mathcal{H}}
\newcommand{\cI}{\mathcal{I}}

\newcommand{\cN}{\mathcal{N}}

\newcommand{\cR}{\mathcal{R}}

\newcommand{\cZ}{\mathcal{Z}}

\newcommand{\kE}{\mathfrak{E}}

\newcommand{\one}{\mathds{1}}
\newcommand{\eps}{\varepsilon}

\DeclareMathOperator{\tr}{tr}

\DeclareMathOperator{\id}{id}

\DeclareMathOperator{\supp}{supp}

\newcommand{\ox}{\otimes}

\newcommand{\Npq}{\cN_{p,\,q}}
\newcommand{\tphi}{\tilde{\phi}}
\DeclareMathOperator{\artanh}{artanh}

\begin{document}
	\title{Dephrasure channel and superadditivity of coherent information}
	\author{Felix Leditzky}\email{felix.leditzky@jila.colorado.edu}
	\affiliation{JILA, University of Colorado/NIST, 440 UCB, Boulder, CO 80309, USA}
	\affiliation{Center for Theory of Quantum Matter, University of Colorado, Boulder, Colorado 80309, USA}
	\author{Debbie Leung}\email{wcleung@uwaterloo.ca}
	\affiliation{Institute for Quantum Computing, University of Waterloo, Waterloo, Ontario, Canada. N2L 3G1.
	}
	\author{Graeme Smith}\email{gsbsmith@gmail.com}
	\affiliation{JILA, University of Colorado/NIST, 440 UCB, Boulder, CO 80309, USA}
	\affiliation{Center for Theory of Quantum Matter, University of Colorado, Boulder, Colorado 80309, USA}
	\affiliation{Department of Physics, University of Colorado, 390 UCB, Boulder, CO 80309, USA}
	\date{\today}
	
	\begin{abstract}
		The quantum capacity of a quantum channel captures its capability for noiseless quantum communication. It lies at the heart of quantum information theory.  Unfortunately, 
		our poor understanding of nonadditivity of coherent information makes it hard to understand the quantum capacity of all but very special channels.  In this paper, we consider the \emph{dephrasure channel}, which is the 
		concatenation of a dephasing channel and an erasure channel.  This very simple channel displays remarkably rich and exotic properties: we find nonadditivity of coherent information at the two-letter level, 
		a big gap between single-letter coherent and private informations, and positive quantum capacity for all complementary channels.  
		Its clean form simplifies the evaluation of coherent information substantially and, as such, we hope that the dephrasure channel will provide a much-needed laboratory for
		the testing of new ideas about nonadditivity.
	\end{abstract}
	
	\maketitle
	
	\section{Introduction}
	A key goal of quantum information theory is to extend the classical theory of information, as pioneered by Shannon \cite{Sha48}, to include quantum effects like superposition and entanglement.  
	The capacity of a noisy communication channel  plays a fundamental role in classical information theory: it is the optimal noiseless communication rate that a noisy channel can support.  
	In the quantum setting, a noisy communication channel has multiple capacities since it can be used to accomplish different communication tasks.  
	Thus, a quantum channel $\cN$ has a capacity for classical communication $C(\cN)$, quantum communication $Q(\cN)$, and private classical communication $P(\cN)$.  
	It is a central challenge of quantum information theory to evaluate these capacities, understand them, and determine their mathematical properties. 
	
	The capacity of a classical channel $\cN\colon X{\;}{\rightarrow}{\;}Y$ is given by $C(\cN){\;}{=}{\;}C^{(1)}(\cN){\;}{=}{\;}\max_{X}I(X;Y)$, where the maximization is over input probability distributions, and the mutual information $I(X;Y){\;}{=}{\;}H(X)+H(Y)-H(XY)$ quantifies the correlations between channel input and output in terms of the Shannon entropy $H(\cdot)$ \cite{Sha48}.  
	This is shown in several steps:
	First, a random-coding argument shows that $C^{(1)}(\cN)$ is an achievable communication rate, so, $C(\cN)\geq C^{(1)}(\cN)$, and $C(\cN)\geq \lim_{n\rightarrow \infty}(1/n)C^{(1)}(\cN^{\otimes n})$.
	Second, Fano's inequality \cite{Fan68} is used to show that $C(\cN) \leq \lim_{n\rightarrow \infty}(1/n)C^{(1)}(\cN^{\otimes n})$, so $C(\cN) = \lim_{n\rightarrow \infty}(1/n)C^{(1)}(\cN^{\otimes n})$. 
	This establishes a multi-letter formula (also called a regularized formula). 
	Third, additivity $C^{(1)}(\cN^{\otimes n}) = n C^{(1)}(\cN)$ is proved to establish the single-letter formula $C(\cN) = C^{(1)}(\cN)$.
	
	Formulas for quantum capacities can be found in a similar way, but for the quantities that are achieved via random coding, additivity in the third step above typically fails.  
	This is fantastic---it means we can achieve higher communication rates than one might naively expect. 
	These rates can be achieved by using error-correcting codes that have more structure than random ensembles. 
	For example, the nonadditivity of the Holevo information $\chi$ shows that entangled signal states can boost the classical capacity of a quantum channel \cite{Has09}, while the nonadditivity of coherent information $I_c$ (defined in \eqref{eq:coh-info}) shows that structured codes can also boost the quantum communication rate over very noisy channels \cite{DSS98}. 
	In the same way, the private information $I_p$ of a quantum channel (defined in \eqref{eq:private-information}) can be nonadditive, in which case the rate of private information transmission is again enhanced by considering structured private codes \cite{SRS08}.
	Quantum information transmission is necessarily private, and hence the private capacity is no less than the quantum capacity.
	However, there are channels showing a strict separation between the two capacities \cite{HHHO05,LLSS14}.
	This property is partly related to nonadditivity issues, as for certain channels with additive coherent and private information such a separation is not possible \cite{Smith08,WatS12}.
	
	The benefits of quantum channels mentioned above also come with frustrations: nonadditivity effects mean that with current techniques, only multi-letter capacity formulas are available for quantum channels. 
	Because these formulas take the form of an optimization over an infinite number of variables, at the moment we have no effective way to evaluate the capacities of a noisy quantum channel.
	
	The main result of this paper is the discovery of a remarkably simple family of quantum channels that display the nonadditivity that makes understanding quantum capacities such a challenge. 
	Dephrasure channels, defined below, show nonadditivity of coherent information in a magnitude that is substantially larger than for the depolarizing channel. 
	Perhaps more importantly, our analysis is much simpler than previous work; this allows for a clearer understanding of the effect.
	Moreover, these dephrasure channels show a strict separation between the coherent information and the private information, strongly suggesting that the respective capacities are strictly separated as well.
	Because of its simple structure and amenability to analysis, we anticipate that the dephrasure channel will become a laboratory for testing new ideas about nonadditivity and quantum channel capacities.
	
	\section{Quantum and private capacity} 
	In quantum information theory, point-to-point communication between a sender and a receiver is modeled by a \emph{quantum channel} $\cN\colon A\to B$, a linear, completely positive, trace-preserving map between the algebras of linear operators of two Hilbert spaces $\cH_A$ and $\cH_B$.
	The quantum capacity $Q(\cN)$ of a quantum channel $\cN$ is defined as the highest rate at which quantum information can be faithfully transmitted through $\cN$ (see App.~\ref{app:entanglement-generation} for an operational definition).
	
	We have the following coding theorem for the quantum capacity \cite{Llo97,BNS98,BKN00,Sho02,Dev05}:
	\begin{equation}
	Q(\cN) = \lim_{n\to\infty} \frac{1}{n} I_c(\cN^{\ox n}) = \sup_{n\in\mathbb{N}}\frac{1}{n} I_c(\cN^{\ox n}),\label{eq:LSD}
	\end{equation}
	where the \emph{channel coherent information} is defined as
	\begin{eqnarray}
	I_c(\cN) &\coloneqq& \max\nolimits_{\rho} I_c(\rho,\cN),\nonumber\\
	\text{with} \quad I_c(\rho,\cN) &\coloneqq& S(\cN(\rho)) - S(\cN^c(\rho)), \label{eq:coh-info}
	\end{eqnarray}
	and $S(\rho)\coloneqq -\tr\rho\log\rho$ is the von Neumann entropy of a state $\rho$ (all logarithms in this paper are taken to base 2).
	In \eqref{eq:coh-info}, $\cN^c\colon A\to E$ denotes a complementary channel of $\cN$, obtained by considering an isometric extension $V\colon \cH_A\to \cH_B\otimes \cH_E$ of $\cN$ satisfying $\cN(\rho) = \tr_E(V\rho V^\dagger)$ \cite{Sti55}, and setting $\cN^c(\rho) \coloneqq \tr_B(V\rho V^\dagger)$.
	
	The optimization in \eqref{eq:LSD} over an (in principle) unbounded number of channel uses $n$ renders the quantum capacity intractable to compute in most cases.
	At the heart of this intractability lies the fact that the coherent information $I_c(\cN)$ can be \emph{superadditive}: there are channels $\cN$ and $n\in\mathbb{N}$ such that $I_c(\cN^{\ox n}) > n I_c(\cN)$.
	A notable example is the qubit depolarizing channel $\cD_q\colon \rho \mapsto (1-q) \rho + q \frac{1}{2}\one$.
	For $q\in [0.2518,0.255]$, it is known that $I_c(\rho_n,\cD_q^{\ox n}) > n I_c(\cD_q)$ for certain input states $\rho_n$ and appropriately chosen $n\geq 3$ \cite{DSS98,SS07,FW08}.
	Moreover, $I_c(\cD_q) = 0$ for $q\geq 0.2524$, such that in the interval $q\in [0.2524,0.255]$ the superadditivity holds in its ``extreme form''.
	There are even more exotic examples of quantum channels exhibiting superadditivity: for any given $n_0\in\mathbb{N}$, there exists a channel $\cN_{n_0}$ such that $I_c(\cN_{n_0}^{\ox n}) = 0$ for all $n\leq n_0$, but the channel still has capacity, $Q(\cN_{n_0})>0$ \cite{CEMOPS15}.
	
	The private capacity $P(\cN)$ of a quantum channel $\cN$ quantifies the optimal rate of transmitting classical data with vanishing probability of error such that the joint environment state of all channel uses has vanishing dependence on the input.
	The private capacity can be expressed as follows \cite{Dev05,CWY04}:
	\begin{equation}
	P(\cN) =  \lim_{n\to\infty} \frac{1}{n} I_p(\cN^{\ox n}) = \sup_{n\in\mathbb{N}}\frac{1}{n} I_p(\cN^{\ox n}),\label{eq:private-capacity}
	\end{equation}
	where the \emph{private information} is defined as
	\begin{equation}
	I_p(\cN) \coloneqq \max_{\kE} I_p(\kE, \cN),\label{eq:private-information}
	\end{equation}
	with the maximization over quantum state ensembles $\kE=\lbrace p_x,\rho_x\rbrace$, and with
	\begin{equation}
	I_p(\kE, \cN) \coloneqq I(X;B)_{\cI \otimes \cN(\rho)}-I(X;E)_{\cI \otimes \cN^c(\rho)}.\label{eq:private-information-ensemble}
	\end{equation}
	The mutual information of a bipartite state $\sigma_{AB}$ is defined as $I(A;B)_\sigma = S(A)_\sigma + S(B)_\sigma - S(AB)_\sigma$, and evaluated in \eqref{eq:private-information-ensemble} on the classical-quantum states $\cI \otimes \cN(\rho_{XA})$ and $\cI \otimes \cN^c(\rho_{XA})$, where $\rho_{XA} = \sum_x p_x |x\rangle\langle x|\ox \rho_x$.
	Quantum information transmission is necessarily private, and hence $P(\cN)\geq Q(\cN)$ for all $\cN$.
	This is also true for the single-letter quantities, $I_p(\cN)\geq I_c(\cN)$.
	
	The private capacity exhibits similarly exotic behavior as the quantum capacity, since the private information defined in \eqref{eq:private-information} is not additive \cite{SRS08,Li09,SS09}.
	Furthermore, there are channels with a large separation of coherent information and private information \cite{LLSS14}.
	
	While the general situation is poorly understood, there are special classes of channels for which the quantum and private capacities can be evaluated.
	A channel $\cN\colon A\to B$ with complementary channel $\cN^c\colon A\to E$ is called \emph{degradable}, if there is another channel $\cD\colon B\to E$ such that $\cN^c = \cD\circ\cN$.
	Degradable channels have additive channel coherent and private informations, $I_c(\cN^{\ox n}) = n I_c(\cN)$ and $I_p(\cN^{\ox n}) = n I_p(\cN)$ for all $n\in\mathbb{N}$, and furthermore they are equal to each other, giving $P(\cN) = Q(\cN) = I_c(\cN) = I_p(\cN)$ for this class of channels \cite{DS05,Smith08}.
	On the other hand, a channel is called \emph{antidegradable}, if there exists a channel $\cA\colon E\to B$ such that $\cN = \cA\circ\cN^c$.
	Due to data processing, antidegradable channels have vanishing channel coherent and private information, and hence $Q(\cN)=0=P(\cN)$ for antidegradable channels.
	
	Generalizing these observations, Watanabe \cite{WatS12} showed that $Q(\cN)=P(\cN)$ if the complementary channel $\cN^c$ has vanishing quantum capacity.
	If furthermore $P(\cN^c)=0$, then all information quantities above are additive, and $I_c(\cN)=I_p(\cN) = Q(\cN)=P(\cN)$ \cite{WatS12}.
	In similar spirit, \cite{CLS17} showed that additivity of coherent information holds for the class of \emph{informationally degradable} channels, which includes all degradable channels \cite{CLS17}.
	Moreover, building on the results in \cite{SSWR15}, we showed in \cite{LLS17} that for a \emph{low-noise channel} $\cN$ that is $\eps$-close to the identity channel in diamond distance, both its quantum and private capacities are within $O(\eps^{3/2}\log \eps)$ of the channel coherent information, limiting the effect of superadditivity for such channels.
	
	Recently, an upper bound on the quantum capacity of a general quantum channel $\cN$ was derived based on a convex decomposition of $\cN$ into degradable and antidegradable maps \cite{LDS17}.
	In the special case of a \emph{flagged} channel 
	\begin{equation}
	\cN = (1-\lambda) \cD \ox |0\rangle\langle 0| + \lambda \cA\ox |1\rangle\langle 1| \label{eq:flagged-channel}
	\end{equation} 
	with $\lambda\in[0,1]$, $\cD$ degradable and $\cA$ antidegradable, optimality of the bound reported in \cite{LDS17} seems intimately connected with whether channels of the form \eqref{eq:flagged-channel} can exhibit superadditivity of coherent information.
	This led us to consider the family of dephrasure channels, which we introduce next.
	
	\section{Dephrasure channel}
	The channel we consider in this paper is composed of dephasing noise followed by erasure, and simply called \emph{dephrasure channel}.
	For two probabilities $p,q\in[0,1]$, it is defined as
	\begin{equation}
	\cN_{p,\,q}(\rho) \coloneqq (1-q) ((1-p) \rho + p Z\rho Z) + q \tr(\rho)|e\rangle\langle e|,\label{eq:dephrasure-channel}
	\end{equation}
	where $Z=|0\rangle\langle 0| - |1\rangle\langle 1|$ is the Pauli $Z$-operator, and $|e\rangle$ is an erasure flag orthogonal to the input space.
	It is not difficult to see that the dephasing channel $\cZ_p\colon \rho\mapsto (1-p) \rho + p Z\rho Z$ is degradable for any $p\in[0,1]$.
	Furthermore, the map $\rho\mapsto  \tr(\rho)|e\rangle\langle e|$ is trivially antidegradable.
	Since $\langle e|\rho|e\rangle=0$ for all qubit input states $\rho$, we can without loss of generality write $\Npq = (1-q) \cZ_p \ox |0\rangle\langle 0| + q\tr(\cdot)|e\rangle\langle e|\ox |1\rangle\langle 1|$, which shows that the dephrasure channel is a flagged channel of the form in \eqref{eq:flagged-channel}.
	
	In the following sections, we analyze the quantum information transmission capabilities of the dephrasure channel.
	Without loss of generality, we restrict the discussion to $p,q\in[0,1/2]$.
	Detailed derivations to all results presented in the sequel can be found in the appendices.
	
	\section{Antidegradability}
	The dephrasure channel $\cN_{p,\,q}$ is degradable only if $q=0$ or if $p=0$ and $q\leq 1/2$.
	For $q\geq 1/2$ the channel is trivially antidegradable due to the antidegradability of the erasure channel $\rho\mapsto (1-q)\rho + q \tr(\rho)|e\rangle\langle e|$ in this range.
	Furthermore, there is a non-trivial region in the $(p,q)$-plane in which $\cN_{p,\,q}$ is antidegradable.
	To determine this region, we consider the following choice of complementary channel:
	\begin{equation}
	\cN^c_{p,\,q}(\rho) \coloneqq q\, \rho \oplus (1-q) \sum_{x=0,\,1} \langle x|\rho|x\rangle |\phi_p^x\rangle\langle \phi_p^x|,\label{eq:complementary-channel}
	\end{equation}
	where $|\phi_p^x\rangle = \sqrt{1-p}\, |0\rangle + (-1)^x \sqrt{p}\,|1\rangle$.
	Making use of unambiguous measurement schemes \cite{Iva87,Die88,Per88}, the original channel $\Npq$ can be recovered from $\Npq^c$ (viz.~$\Npq^c$ can be degraded to $\Npq$) in the region
	\begin{eqnarray}
	\cA &\coloneqq& \lbrace (p,q)\colon p\in[0,1/2],q\geq k(p) \rbrace, \label{eq:A}\\
	k(p) &\coloneqq& \frac{1-2p}{2(1-p)}\label{eq:k}\,.
	\end{eqnarray}
	We refer to App.~\ref{app:antidegradability} for details of this calculation.
	
	\section{Single-letter coherent information} 
	In order to analyze nonadditivity properties of the dephrasure channel, we first derive a formula for the single-letter coherent information $I_c(\Npq)$ defined in \eqref{eq:coh-info}.
	
	The dephrasure channel is defined in terms of a $Z$-dephasing, and therefore one could expect that the coherent information $I_c(\rho,\Npq)$ in \eqref{eq:coh-info} is maximized by states $\rho_z = (0,0,z)$ that are diagonal in the $Z$-eigenbasis and hence invariant under $Z$-dephasing.
	Indeed, ordinary calculus shows (see App.~\ref{app:single-letter-ci}) that $I_c(\cN_{p,\,q}) = \max_{\rho_z} I_c(\rho_z,\Npq)$ in the region 
	\begin{eqnarray}
	\cR_1 &\coloneqq& \left\lbrace (p,q)\colon p\in[0,1/2],\, 0\leq q < g(p) \right\rbrace,\label{eq:R1}\\
	g(p) &\coloneqq& \frac{(1-2p)^2}{1+(1-2p)^2}.\label{eq:g}
	\end{eqnarray} 
	Numerics show that $\cR_1$ also includes the region where $I_c(\cN_{p,\,q}) \geq 0$ (see Fig.~\ref{fig:cohinfo-n1}).
	For states $\rho_z=(0,0,z)$, we have the explicit formula
	\begin{equation}
	I_c(\rho_z,\Npq) = (1-2q)S(\rho_z) - (1-q)S(\Phi_{p,z})\label{eq:cho-info-z}
	\end{equation} 
	with
	\begin{equation} 
	\Phi_{p,\,z} = \begin{psmallmatrix}
	1-p & z \sqrt{p(1-p)} \\ z\sqrt{p(1-p)} & p
	\end{psmallmatrix}.\nonumber
	\end{equation}
	We prove in App.~\ref{app:threshold} that $I_c(\rho_z,\Npq)>0$ if and only if $(p,q)\in\cR_1$.
	
	The formula \eqref{eq:cho-info-z} has a maximum at $z=0$ in the region
	\begin{eqnarray}
	\cR_2 &\coloneqq& \left\lbrace (p,q)\colon p\in[0,1/2],\, 0 \leq q < j(p)\right\rbrace\label{eq:R2}\\
	j(p) &\coloneqq& \frac{1-2p-2p(1-p)\ln\frac{1-p}{p}}{2-4p-2p(1-p)\ln\frac{1-p}{p}},\label{eq:j}
	\end{eqnarray}
	that is, in this region the completely mixed state $\pi=\frac{1}{2}\one$ maximizes the coherent information, which evaluates to $I_c(\cN_{p,\,q}) = I_c(\pi,\cN_{p,\,q}) = 1-2q-(1-q)h(p)$.
	
	To sum up, in the region $\cR_1$ defined in \eqref{eq:R1} the coherent information $I_c(\Npq)$ is maximized by states diagonal in the $Z$ eigenbasis. 
	In the subregion $\cR_2\subseteq\cR_1$ defined in \eqref{eq:R2}, the coherent information $I_c(\Npq)$ is maximized by the completely mixed state $\pi$, and evaluates to $I_c(\pi,\Npq) = 1-2q-(1-q)h(p).$
	In the fish-shaped region $\cF \coloneqq \cR_1\setminus\cR_2 = \left\lbrace (p,q)\colon p\in[0,1/2],\, j(p)< q <g(p) \right\rbrace,$
	the coherent information is maximized by $Z$-diagonal states with $z\neq 0$.
	Furthermore, the $0$-contour line of $I_c(\Npq)$ coincides with $g(p)$. 
	Fig.~\ref{fig:repetition-code-n2} plots the functions $g(p)$, $j(p)$ and $k(p)$ that bound these regions.
	
	\section{Superadditivity of coherent information}
	
	\begin{figure}[t]
		\centering
		\includegraphics[width=\columnwidth]{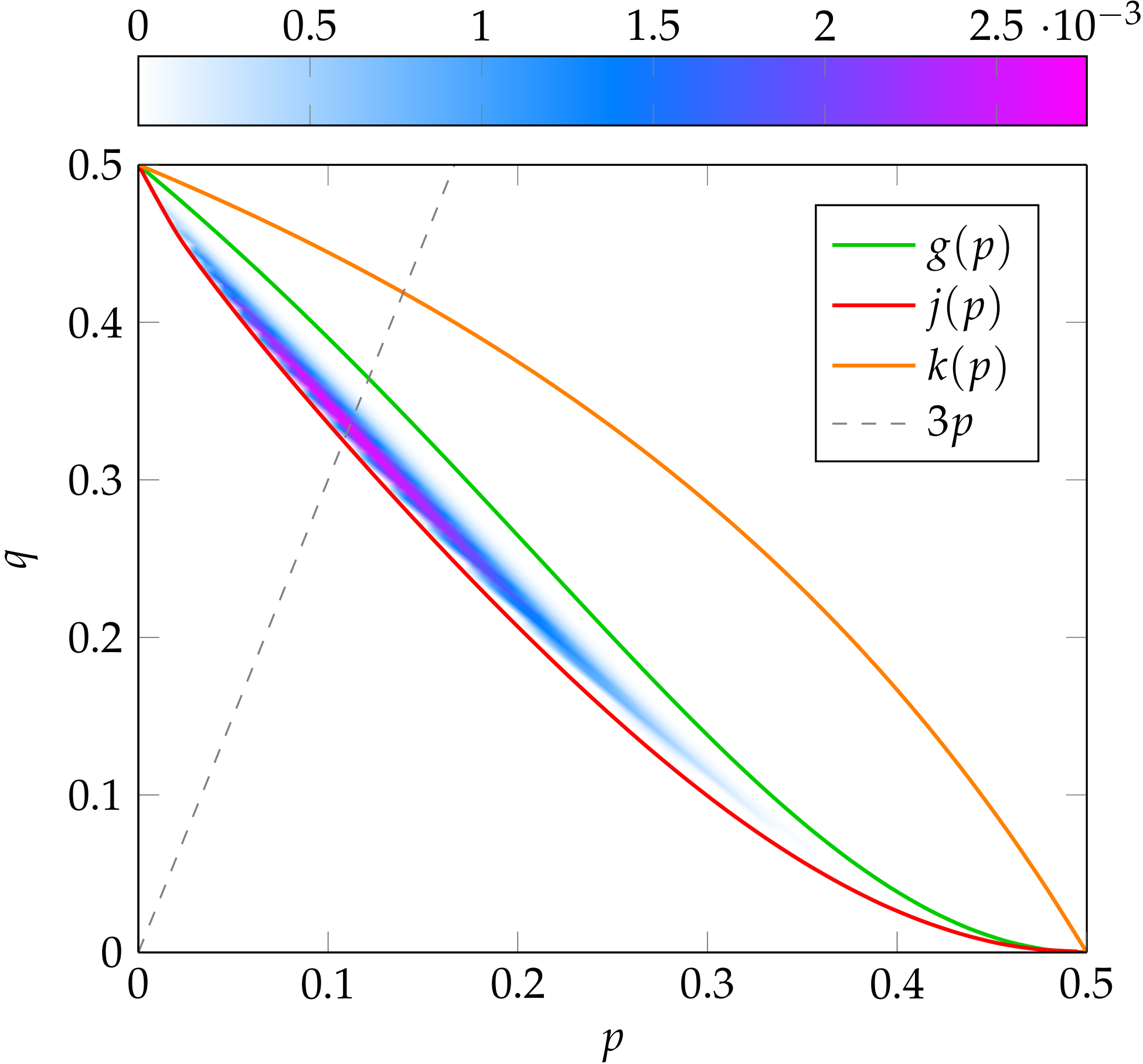}
		\caption{Heat map of the quantity $\max_\lambda \frac{1}{2} I_c(\rho_2,\Npq^{\ox 2}) - I_c(\Npq)$. The repetition code $\rho_2$ is defined in \eqref{eq:repetition-code}.
			The functions $g(p)$ (green) and $j(p)$ (red) are defined in \eqref{eq:g} and \eqref{eq:j}, respectively.
			The function $k(p)$ (orange) defined in \eqref{eq:k} bounds the region $\cA$ of antidegradability of $\Npq$ defined in \eqref{eq:A}.}
		\label{fig:repetition-code-n2}
	\end{figure}

\begin{figure}[t]
	\centering
	\includegraphics{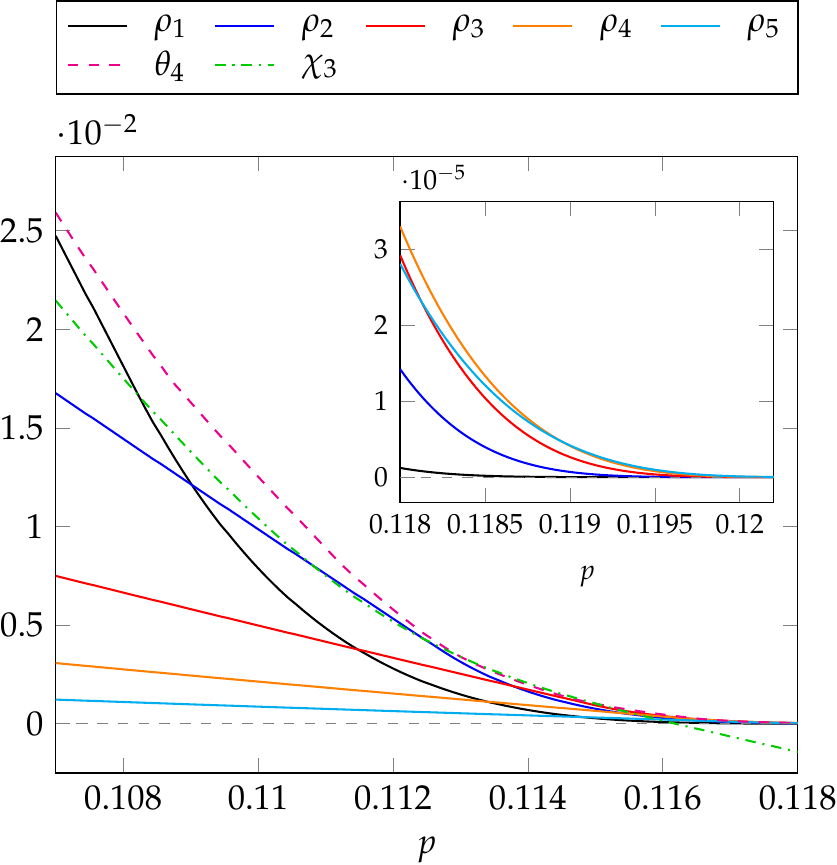}
	
	\caption{Plot of the coherent information $I_c(\cdot,\cN_{p,\,3p}^{\ox n})/n$ in the interval $p\in[0.107,0.118]$ for the repetition code $\rho_n$ for $n=1,\dots,5$ defined in \eqref{eq:repetition-code} (solid lines), the generalized $Z$-diagonal code $\theta_4$ defined in \eqref{eq:dephased-code} in App.~\ref{app:superadditivity} for $k=4$ (dashed line), and the non-diagonal code $\chi_3$ defined in \eqref{eq:chi3} in App.~\ref{app:superadditivity} (dash-dotted lines). 
		The zero line is plotted as a dashed gray line for reference.
		The inset plot shows the repetition codes $\rho_n$ in the interval $p\in[0.118,0.1202]$, where the single-letter coherent information becomes vanishingly small, while repetition codes for $n\geq 2$ still give a substantial contribution.
	}
	\label{fig:repcodes}
\end{figure}
	
	In this section we show that the dephrasure channel $\Npq$ exhibits superadditivity of the coherent information within the region $\cF$: there are $(p,q)\in\cF$ for which $I_c(\Npq^{\ox n})> n I_c(\Npq)$.
	
	We first demonstrate superadditivity of $I_c(\Npq)$ using a simple (weighted) $n$-repetition code 
	\begin{equation}
	\rho_n \coloneqq \lambda |0\rangle\langle 0|^{\ox n} + (1-\lambda) |1\rangle\langle 1|^{\ox n},
	\label{eq:repetition-code}
	\end{equation}
	where $\lambda\in[0,1]$.
	Observe that
	\begin{equation}
	I_c(\rho_n,\Npq^{\ox n}) = S\left(\Npq^{\ox n}(\rho_n)\right) - S\left(\Npq^{\ox n}(\phi_n)\right),\label{eq:coh-info-purification}
	\end{equation}
	where in the second term, the output entropy of the complementary channel in (\ref{eq:coh-info}) is rephrased in terms of the entropy of the purification of $\rho_n$, $\phi_n\equiv |\phi_n\rangle\langle\phi_n|$ with $|\phi_n\rangle \coloneqq \sqrt{\lambda} |0\rangle^{\ox n+1} + \sqrt{1-\lambda} |1\rangle^{\ox n+1}$, and $\Npq^{\ox n}$ acts on all but the first (purifying) system of $\phi_n$.

	The expression \eqref{eq:coh-info-purification} is independent of the particular purification of $\rho_n$.
	To evaluate it, note that $\Npq^{\ox n}$ is a sum of channels involving $i$ erasures and $n-i$ dephasing errors for $i=0,\dots,n$.
	Any two erasure patterns differing in at least one position yield orthogonal output states, and hence \eqref{eq:coh-info-purification} splits up into a sum over the different erasure patterns.
	Moreover, for a fixed erasure pattern with $1\leq i\leq n-1$ erasures the two entropy terms on the right-hand side of \eqref{eq:coh-info-purification} yield the same value $h(\lambda)\coloneqq-\lambda\log\lambda-(1-\lambda)\log(1-\lambda)$, the binary entropy of $\lambda$.
	Hence, we only need to evaluate \eqref{eq:coh-info-purification} in the cases of $n$ dephasing erorrs and $n$ erasures.
	Our calculation in App.~\ref{app:superadditivity} yields
	\begin{eqnarray}
	&& \hspace{-4ex} I_c(\rho_n,\Npq^{\ox n}) = ((1-q)^n-q^n) \; h(\lambda)\nonumber\\
	&& 
	\hspace{-2ex}
	-(1-q)^n\left(1- \frac{u}{2}\log\frac{1+u}{1-u}
	- \frac{1}{2}\log\left(1-u^2\right)\right)\!,
	\label{eq:coh-info-repetition-code}
	\end{eqnarray}
	for $u = u(\lambda,p,n) = \sqrt{1- 4\lambda(1-\lambda)(1-(1-2p)^{2n})}$.
	
	The formula \eqref{eq:coh-info-repetition-code} provides examples of superadditivity of the coherent information of $\Npq$.
	This is demonstrated in Fig.~\ref{fig:repetition-code-n2}, where we plot a heat map of the quantity $
	\max_\lambda \frac{1}{2} I_c(\rho_2,\Npq^{\ox 2}) - I_c(\Npq)$.
	The region with the largest values of this quantity is colored in purple in Fig.~\ref{fig:repetition-code-n2}, and crossed by the $(p,3p)$-diagonal (dashed line).
	We therefore further investigate the optimized coherent information of the repetition code \eqref{eq:repetition-code} along this diagonal for $n=1,\dots, 5$ \footnote{Note that the choice of the particular diagonal $(p,3p)$ is only exemplary; the effects of superadditivity of coherent information, as well as the separation between coherent information and private information occur along any diagonal.}.
	In Fig.~\ref{fig:repcodes}, we plot $\max_{\lambda} I_c(\rho_n,\cN_{p,\,3p}^{\ox n})/n$ for $1\leq n\leq 5$ in the intervals $p\in[0.107,0.118]$ and $p\in[0.118,0.1202]$.
	In the latter interval, the single-letter coherent information becomes vanishingly small, while repetition codes for $n\geq 2$ still give a substantial contribution.
	However, we prove in App.~\ref{app:threshold} that the weighted repetition code has the same threshold given by $g(p)$ for all $n\in\mathbb{N}$, which includes the optimal single-letter code for $n=1$.
	
	We were also able to find codes that outperform the weighted repetition code \eqref{eq:repetition-code}.
	These include more general $Z$-diagonal codes (such as $\theta_4$ in Fig.~\ref{fig:repcodes}, defined in \eqref{eq:dephased-code} in App.~\ref{app:superadditivity}, as well as certain \emph{non-diagonal} codes (such as $\chi_3$ in Fig.~\ref{fig:repcodes}, defined in \eqref{eq:chi3} in App.~\ref{app:superadditivity}.
	Furthermore, other interesting non-diagonal codes can be found using a neural network state ansatz \cite{BL18}.

	\section{Separation of private information and coherent information}
	
		\begin{figure}[t]
		\centering
		\includegraphics[]{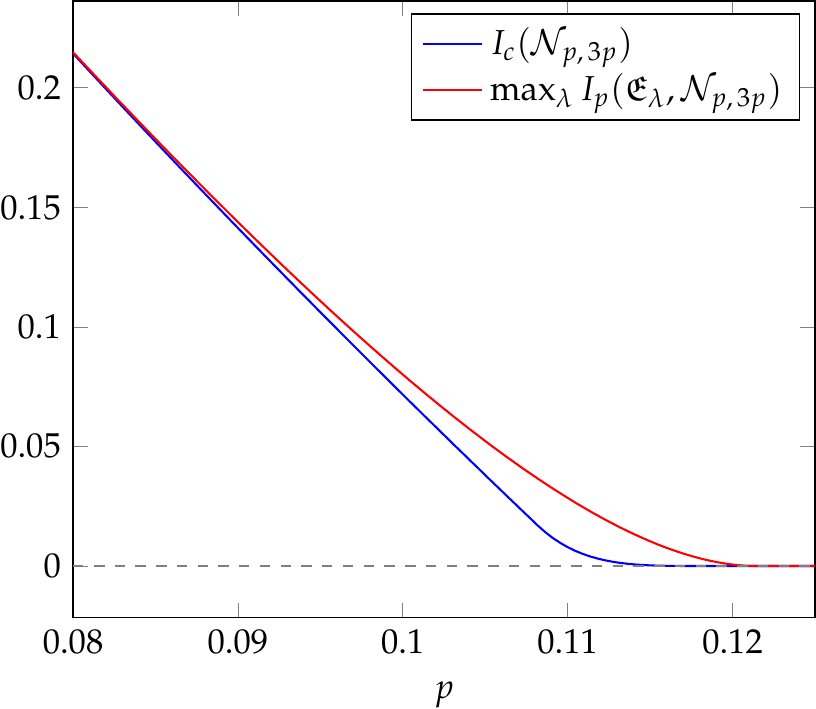}
		\caption{Plot of the optimal single-letter coherent information $I_c(\cN_{p,\,3p})$ for $p\in[0.08,0.125]$ (blue), and a lower bound to the single-letter private information, $\max_\lambda I_p(\kE_\lambda,\cN_{p,\,3p})$ (red), where the private code $\kE_\lambda$ is defined in \eqref{eq:opt-private-ensemble}.}
		\label{fig:private-information}
	\end{figure}
	
	Finally, we investigate the capabilities of private information transmission of the dephrasure channel.
	Numerical investigations (see App.~\ref{app:separation}) suggest that the following ensemble $\kE_\lambda = \lbrace p_1,\rho_1; p_2,\rho_2\rbrace$
	maximizes the single-letter private information $I_p(\cN_{p,\,3p})$:
	\vspace*{-0.1cm}
	\begin{eqnarray}
	p_1 &=& \tfrac{1}{2}, \quad \rho_1 = \lambda |+\rangle\langle +| + (1-\lambda) |-\rangle\langle -|\nonumber\\
	p_2 &=& \tfrac{1}{2}, \quad \rho_2 = (1-\lambda) |+\rangle\langle +| + \lambda |-\rangle\langle -|, \label{eq:opt-private-ensemble}\\
	&& \nonumber
	\end{eqnarray}
	where $\lambda\in[0,1]$ and $|\pm\rangle = (|0\rangle \pm |1\rangle)/\sqrt{2}$.
	This private code shows a strict separation between $I_c(\cN_{p,\,3p})$ and $I_p(\cN_{p,\,3p})$ in the interval $p\in[0.08,0.125]$, plotted in Fig.~\ref{fig:private-information}.
	We note that the private information remains positive up to $p\simeq 0.12145$, which is exactly where the diagonal $(p,3p)$ meets the curve $g(p)$ defined in \eqref{eq:g} marking the threshold of $I_c(\cN_{p,\,q})$.
	
	It is an interesting open question whether the dephrasure channel also exhibits superadditivity of private information.
	However, to demonstrate this effect one first needs to determine the optimal single-letter private information.
	We conjecture the private code in \eqref{eq:opt-private-ensemble} to be optimal for the dephrasure channel $\cN_{p,\,3p}$.
	
	Finally, we note that the complementary channel $\Npq^c$ has positive coherent information for all $p,q\in(0,1/2]$ (see App.~\ref{app:coh-info-complementary}), which implies that $P(\Npq^c)\geq Q(\Npq^c)>0$ for all $p,q\in(0,1/2]$.
	Similarly as for the depolarizing channel \cite{LW17}, this indicates that Watanabe's results \cite{WatS12} cannot be applied to the dephrasure channel. 
	
	\begin{acknowledgments}
		The authors are grateful to Vikesh Siddhu for pointing out to us that the single-letter coherent information remains non-zero up to $g(p)$ and other helpful comments.
		DL is supported by an NSERC discovery grant and a CIFAR research grant via the Quantum Information Science program. FL and GS are supported by National Science Foundation (NSF) Grant No.~PHY 1734006. GS is supported by the NSF Grant No.~CCF 1652560.
		The authors appreciate the hospitality of JILA at the University of Colorado Boulder, the Institute for Quantum Computing at the University of Waterloo, and the Kavli Institute for Theoretical Physics at the University of California Santa Barbara, where parts of this work were completed.
	\end{acknowledgments}

\onecolumngrid
\appendix

\section{Preliminaries}
Let $\cB(\cH)$ denote the set of linear operators on a Hilbert space $\cH$.
A state $\rho\in\cB(\cH)$ is a positive semidefinite linear operator with unit trace.
The \emph{von Neumann entropy} $S(\rho)$ of a state $\rho$ is defined as $S(\rho)= -\tr\rho\log\rho$.
In this paper, $\log$ and $\ln$ denote the logarithms of base $2$ and $e$, respectively.
For a state $\rho$ with spectral decomposition $\rho=\sum_i \lambda_i \pi_i$, the support of $\rho$ is defined as $\supp\rho \coloneqq \sum_{i\colon \lambda_i >0} \pi_i$.
If $\rho = \sum_i p_i \rho_i$ with $\supp\rho_i\perp \supp\rho_j$ for $i\neq j$, then 
\begin{align}
S(\rho) = H(\lbrace p_i\rbrace) + \sum_i p_i S(\rho_i),
\label{eq:entropy-cq-factorization}
\end{align} where $H(\lbrace p_i\rbrace) \coloneqq -\sum_i p_i\log p_i$ is the \emph{Shannon entropy} of the probability distribution $\lbrace p_i\rbrace$.
For a binary distribution $\lbrace p,1-p\rbrace$ with $p\in[0,1]$, we denote by $h(p)\coloneqq H(\lbrace p,1-p\rbrace)$ the \emph{binary entropy} of $p$.

\section{Entanglement generation and quantum capacity}\label{app:entanglement-generation}

The capability of a quantum channel $\cN$ to faithfully transmit quantum information from $A$ to $B$ is quantified by its \emph{quantum capacity} $Q(\cN)$ \cite{Sch96,SN96,Llo97,BNS98,BKN00,Sho02,Dev05}, which can be defined in the following way \cite{Dev05}.

Suppose that Alice sends the $A^n$ part of a pure state $|\psi\rangle_{RA^n}$ through $n$ copies of the channel $\cN$ to Bob, who applies a decoding quantum operation $\cD_n\colon B^n\to R'$ to obtain a state $\sigma_{RR'} = (\id_R\ox \cD_n\circ \cN^{\ox n})(\psi_{RA^n})$. 
If $\sigma_{RR'}$ approaches $m_n$ copies of a maximally entangled state $(|00\rangle + |11\rangle)/\sqrt{2}$ as $n\to\infty$ with respect to a suitable distance measure (such as fidelity), we say that $\lim_{n\to\infty} m_n/n$ is an \emph{achievable rate} for quantum information transmission through $\cN$.
The quantum capacity $Q(\cN)$ is defined as the supremum over all achievable rates.

\section{Antidegradability of the dephrasure channel}\label{app:antidegradability}
The channel $\cN_{p,\,q}$ is a convex mixture of a degradable channel (the dephasing channel $\cZ_p$) and an antidegradable channel (the completely depolarizing channel $\rho\mapsto \tr(\rho)|e\rangle\langle e|$).
We consider the following choice of complementary channel $\cN_{p,\,q}^c$,
\begin{equation}
\cN^c_{p,\,q}(\rho) \coloneqq q\, \rho \oplus (1-q) \sum_{x=0,\,1} \langle x|\rho|x\rangle |\phi_p^x\rangle\langle \phi_p^x|,\label{eq:complementary-channel-app}
\end{equation}
where $|\phi_p^x\rangle = \sqrt{1-p}\, |0\rangle + (-1)^x \sqrt{p}\,|1\rangle$.
It is easy to see that $\Npq$ is antidegradable for $q\in[1/2,1]$ and any $p\in[0,1]$, since the map
\begin{align}
\widetilde{\cA}_{p,\,q} \coloneqq \cE_{(2q-1)/q}\circ\cZ_p \ox \tr(\cdot |0\rangle\langle 0|) + \cE_1 \ox \tr(\cdot |1\rangle\langle 1|),\label{eq:trivial-antidegrading-map}
\end{align}
satisfies $\cN_{p,\,q} = \widetilde{\cA}_{p,\,q}\circ\cN_{p,\,q}^c$, and is CP for $q\geq 1/2$.

\begin{figure}[t]
	\centering
	\includegraphics[width=0.7\textwidth]{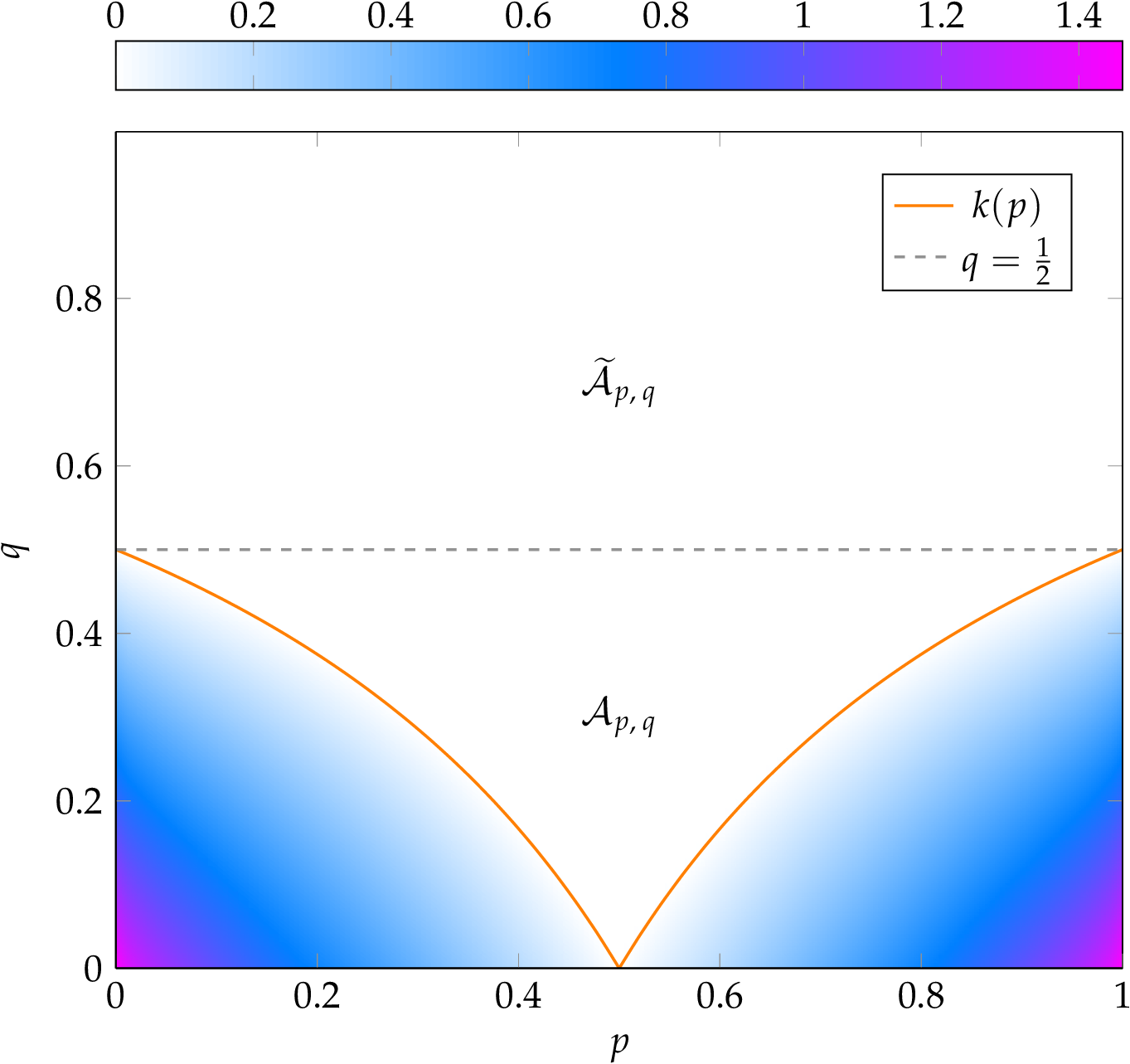}
	\caption{Plot of the antidegradability parameter $\operatorname{adg}(\Npq)$ from \cite{SSWR15} for $p,q\in[0,1]$. The channel is antidegradable if $\operatorname{adg}(\Npq)=0$, which corresponds to the white region in the plot.
		In the region $\lbrace (p,q)\colon p\in[0,1],\, q\in[1/2,1]\rbrace$ above the dashed gray line $q=1/2$, the `trivial' map $\widetilde{\cA}_{p,\,q}$ defined in \eqref{eq:trivial-antidegrading-map} antidegrades $\Npq$.
		In the region $\lbrace (p,q)\colon p\in[0,1],\, q\in[k(p), 1/2]\rbrace$ between the dashed gray line and the orange line corresponding to the function $k(p)$ defined in \eqref{eq:k-app}, the `non-trivial' map $\cA_{p,\,q}$ defined in \eqref{eq:non-trivial-antidegrading-map} antidegrades $\Npq$.}
	\label{fig:antideg}
\end{figure}

However, numerically investigating antidegradability of $\Npq$ (e.g., by solving the corresponding semidefinite program in \cite{SSWR15}) shows that, in fact, $\Npq$ is also antidegradable for certain $(p,q)$ with $q<1/2$, as can be seen in Fig.~\ref{fig:antideg}.
Indeed, for $p\in[0,1/2]$ consider the following map:
\begin{align}
\cA_{p,\,q} \coloneqq{} & \cE_x \ox \tr(\cdot |0\rangle\langle 0|) + \sum_{i=0,\,1,\,e} \tr(\cdot\Pi_i) |i\rangle\langle i| \ox \tr(\cdot |1\rangle\langle 1|),\label{eq:non-trivial-antidegrading-map}
\end{align}
where $x = 1 - \frac{(1-q)(1-2p)}{q}$, and the POVM $\lbrace \Pi_0,\Pi_1,\Pi_e\rbrace$ has the effect operators
\begin{align}
\begin{aligned}
\Pi_0 = {} & \frac{1}{2(1-p)}\begin{pmatrix} p & \sqrt{p(1-p)} \\ \sqrt{p(1-p)} & 1-p \end{pmatrix}\\
\Pi_1 = {} & \frac{1}{2(1-p)}\begin{pmatrix} p & -\sqrt{p(1-p)} \\ -\sqrt{p(1-p)} & 1-p \end{pmatrix}\\
\Pi_e = {} & \frac{1-2p}{1-p} \begin{pmatrix} 1 & 0\\ 0 & 0 \end{pmatrix}.
\end{aligned}
\label{eq:POVM}
\end{align}
The map $\cA_{p,\,q}$ defined in \eqref{eq:non-trivial-antidegrading-map} satisfies $\Npq = \cA_{p,\,q}\circ \Npq^c$.
Moreover, $\cA_{p,\,q}$ is CP for $x\geq 0$, which is equivalent to 
\begin{align}
q\geq k(p)\coloneqq \begin{cases} \dfrac{1-2p}{2(1-p)} & \text{if $p\in[0,1/2]$,}\\[1.5em] \dfrac{1-2(1-p)}{2p} & \text{if $p\in[1/2,1]$}.
\end{cases}\label{eq:k-app}
\end{align}
Hence, for $(p,q)\in[0,1/2]$ the region of antidegradability is
\begin{align}
\cA &\coloneqq \lbrace (p,q)\colon p\in[0,1/2],q\geq k(p) \rbrace \label{eq:A-app}.
\end{align}

The map $\cA_{p,\,q}$ defined in \eqref{eq:non-trivial-antidegrading-map} is based on an unambiguous measurement scheme:
In unambiguous state discrimination, the task is to distinguish between two non-orthogonal pure states $\psi_0$ and $\psi_1$ with a three-outcome measurement with effect operators $\lbrace \Pi_0, \Pi_1, \Pi_?\rbrace$ in such a way that an outcome `$0$' or `$1$' always yields the right answer, i.e., $\langle\psi_0| \Pi_1|\psi_0\rangle = 0 = \langle \psi_1| \Pi_0 | \psi_1\rangle$.
However, due to $\langle\psi_0|\psi_1\rangle \neq 0$, it necessarily holds that $\Pi_0+\Pi_1 \neq \one$, and the completing effect operator $\Pi_?=\one - \Pi_0 - \Pi_1$ corresponds to an inconclusive outcome.
If the discrimination is \emph{unbiased}, i.e., the prior probabilities of $\psi_i$ are each equal to $\frac{1}{2}$, then \textcite{Iva87,Die88,Per88} showed that the minimal probability $P(?)$ of obtaining an inconclusive measurement outcome is given by
\begin{align}
P(?)=\frac{1}{2} ( \langle \psi_0|\Pi_?|\psi_0\rangle + \langle \psi_1|\Pi_?|\psi_1\rangle) = |\langle \psi_0|\psi_1\rangle |.\label{eq:min-inconclusive-prob}
\end{align}

Considering the complementary channel $\Npq^c$ defined in \eqref{eq:complementary-channel-app}, successfully distinguishing between the states $\phi_p^0$ and $\phi_p^1$ (and thus inferring $\langle x|\rho|x\rangle$ from $\sum_x \langle x|\rho|x\rangle \phi_p^x$) clearly helps in the attempt to degrade $\Npq^c$ to $\Npq$. 
In the event of an inconclusive outcome `$?$', erasure is the only reasonable option (hence the labeling of $\Pi_e$).
The POVM given in \eqref{eq:POVM} was derived in \cite{Iva87,Die88,Per88} and achieves \eqref{eq:min-inconclusive-prob}.

\section{Maximizing the single-letter coherent information}\label{app:single-letter-ci}
In the following, we restrict our discussion of the dephrasure channel $\cN_{p,\,q}$ to the region $\lbrace (p,q)\colon 0\leq p,q,\leq 1/2\rbrace$.
Let $(x,y,z)$ with $x,y,z\in[-1,1]$ and $\|(x,y,z)\|\leq 1$ be the Bloch vector of a qubit state $\rho$, defined via the Bloch representation $\rho = \frac{1}{2}(\one + x X + y Y + z Z)$.
Furthermore, let $U$ be the rotation in the Bloch sphere that rotates a Bloch vector into the $xz$-plane, i.e., $U(x,y,z) \mapsto (x',0,z)$ where $x'^2 = x^2+y^2$.
It is easy to see that $\Npq$ is covariant with respect to $U$, and the same holds for the complementary channel $\Npq^c$ defined in \eqref{eq:complementary-channel-app}.
Hence, without loss of generality we can restrict to states of the form $(x,0,z)$ in the maximization of $I_c(\Npq)$.

Since $\Npq(\rho) = (1-q) \cZ_p(\rho) + q|e\rangle\langle e|$ and $|e\rangle \in (\supp \rho_p)^\perp$, we have $S(\Npq(x,0,z)) = h(q) + (1-q)S(\cZ_p(\rho))$.
Furthermore, $\Npq^c(\rho) = q\,\rho\ox |0\rangle\langle 0| + (1-q) \Phi_{p,\,z} \ox |1\rangle\langle 1|$, where 
\begin{align}
\Phi_{p,\,z} = \begin{pmatrix}
1-p & z \sqrt{p(1-p)} \\ z\sqrt{p(1-p)} & p
\end{pmatrix}.
\end{align}
Hence, $S(\Npq^c(\rho)) = h(q) + q S(\rho) + (1-q) S(\Phi_{p,\,z})$.
In summary,
\begin{align}
I_c(\rho,\Npq) &= S(\Npq(\rho)) - S(\Npq^c(\rho))\\
&= (1-q) S(\cZ_p(\rho)) - q S(\rho) - (1-q)S(\Phi_{p,\,z}).\label{eq:coh-info-xz}
\end{align}

\begin{figure}[t]
	\centering
	\includegraphics[width=0.7\textwidth]{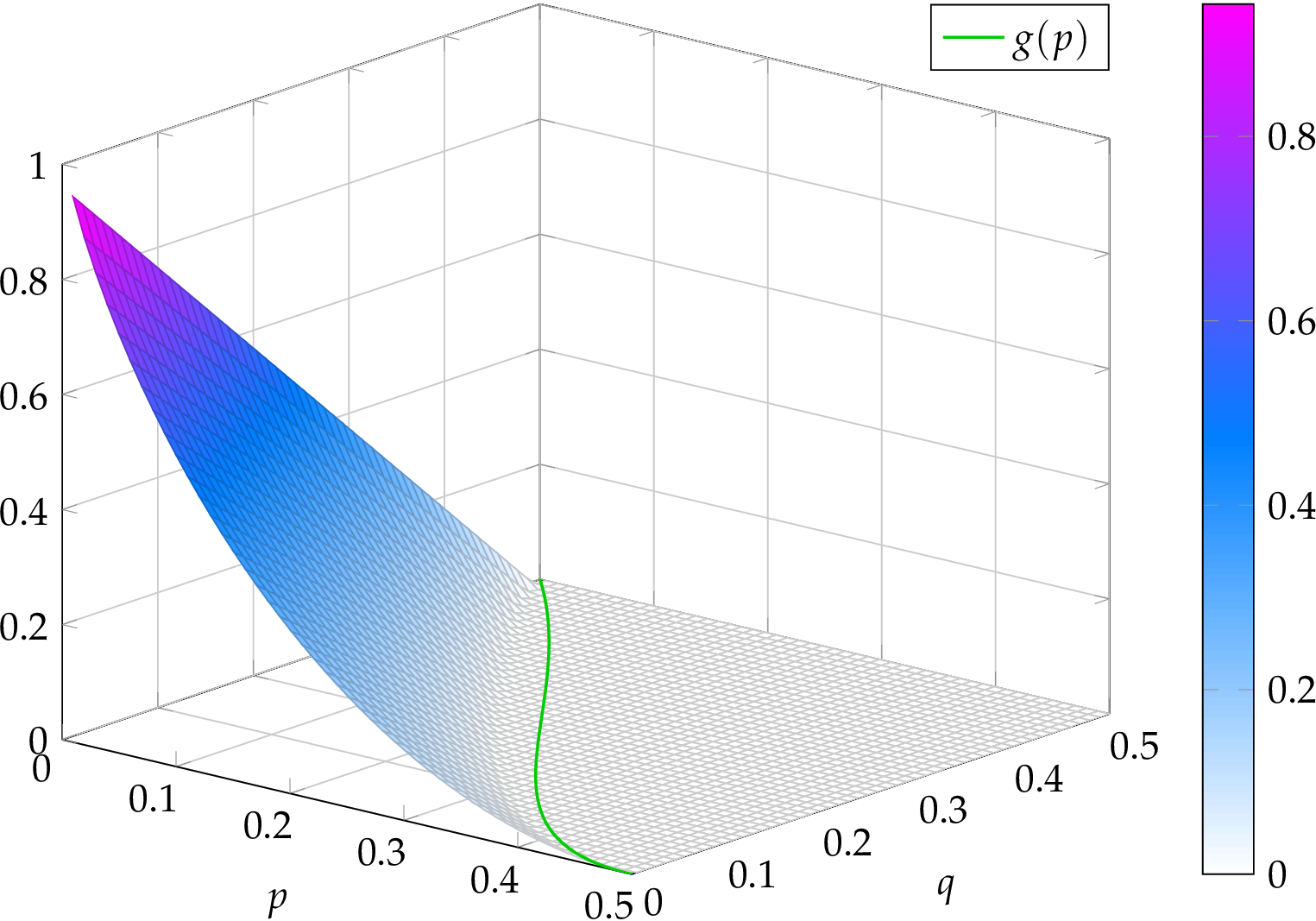}
	\caption{Plot of the coherent information $I_c(\Npq)$ for $p,q\in[0,1/2]$. The function $g(p)$ (green) is defined in \eqref{eq:g-app}.}
	\label{fig:cohinfo-n1}
\end{figure}

In the following we show that whenever $I_c(\rho,\Npq)$ as given in \eqref{eq:coh-info-xz} is non-negative, it is maximized by states that are diagonal in the $Z$-eigenbasis.
Setting $f(x)\coloneqq (1-q) S(\cZ_p(\rho))-qS(\rho)- (1-q)S(\Phi_{p,\,z})$ and noting that $\Phi_{p,\,z}$ is independent of $x$, we calculate $f'(x)$:
\begin{multline}
f'(x) = -\frac{2x}{4(1-2p)(x^2+z^2)} \left(q(1-2p)\sqrt{x^2+z^2}\log\frac{1-\sqrt{x^2+z^2}}{1+\sqrt{x^2+z^2}} \right.\\
\left. - (1-q)(1-2p)^2 \sqrt{x^2+z^2}\log\frac{1-\sqrt{(1-2p)^2x^2+z^2}}{1+\sqrt{(1-2p)^2x^2+z^2}}\right)
\end{multline}
Assuming that $x^2+z^2>0$ (since otherwise there is nothing to show), we see that $x=0$ is a critical point of $f(x)$, with second derivative equal to
\begin{align}
f''(x)\Bigr|_{x=0} = \frac{\log\frac{1-|z|}{1+|z|}}{2|z|} (1-2q-4p(1-p)(1-q)).
\end{align}
Note that $\frac{1}{2|z|}\log\frac{1-|z|}{1+|z|}$ is defined for $z=0$, and $\frac{1}{2|z|}\log\frac{1-|z|}{1+|z|}<0$ for any $z\in(-1,1)$.
Hence, $f''(0) < 0$, and therefore $f(x)$ attains a maximum at $x=0$ if 
\begin{align}
q < g(p)\coloneqq \frac{(1-2p)^2}{1+(1-2p)^2}.\label{eq:g-app}
\end{align}
Moreover, numerics demonstrate that the region 
\begin{align}
\cR_1\coloneqq \left\lbrace (p,q)\colon p\in[0,1/2],\, 0\leq q < g(p) \right\rbrace \label{eq:R1-app}
\end{align} 
contains the region of non-negative coherent information (see Fig.~\ref{fig:cohinfo-n1}).
Therefore, it suffices to only consider states of the form $(0,0,z)$ in the maximization of the coherent information given in \eqref{eq:coh-info-xz}.
For these states, the formula \eqref{eq:coh-info-xz} reduces to
\begin{align}
I_c(\rho_z,\Npq) &= (1-2q)S(\rho_z) - (1-q)S(\Phi_{p,z})\\
&\eqqcolon i_c(z).
\end{align}
We prove in App.~\ref{app:threshold} that the threshold $of I_c(\rho_z,\Npq)$ in the $(p,q)$-plane coincides with the function $g(p)$ defined in \eqref{eq:g-app}.

\begin{figure}[t]
	\centering
	\begin{tikzpicture}
	\begin{axis}[
	domain = 0:0.5,
	xlabel = $p$,
	ylabel = $q$,
	xmin = 0,
	ymin = 0,
	xmax = 0.5,
	ymax = 0.5
	]
	\addplot[name path=plot1,smooth,thick,color=dgreen] {(1-2*x)^2/(1+(1-2*x)^2)};
	\addplot[name path=plot2,smooth,thick,color=red,domain=0:0.495] {(1-2*x-2*x*(1-x)*ln((1-x)/x))/(2-4*x-2*x*(1-x)*ln((1-x)/x))};
	\addplot[smooth,thick,color=red,domain=0.495:0.5] {8/3*(x-0.5)^2};
	\addplot[smooth,color= orange, thick] {(1-2*x)/(2*(1-x))};
	\addplot[blue!10] fill between[of = plot1 and plot2];
	\legend{$g(p)$,$j(p)$,,$k(p)$};
	\node[color=blue!80,rotate=-45] at (axis cs:0.18,0.18) {$\cF=\cR_1\setminus\cR_2$};
	\node[color=orange] at (axis cs:0.3,0.4) {$\cA$};
	\end{axis}
	\end{tikzpicture}
	\caption{Plot of the fish-shaped region $\cF=\cR_1\setminus\cR_2$ (blue) defined in \eqref{eq:F}, where $\cR_1$ is defined in \eqref{eq:R1-app} and bounded by $g(p)$ (green) as given in \eqref{eq:g-app}, and $\cR_2$ is defined in \eqref{eq:R2-app} and bounded by $j(p)$ (red) as given in \eqref{eq:j-app}.
		The function $k(p)$ (orange) defined in \eqref{eq:k-app} bounds the region $\cA$ of antidegradability of $\Npq$ defined in \eqref{eq:A-app}.}
	\label{fig:fish}
\end{figure}
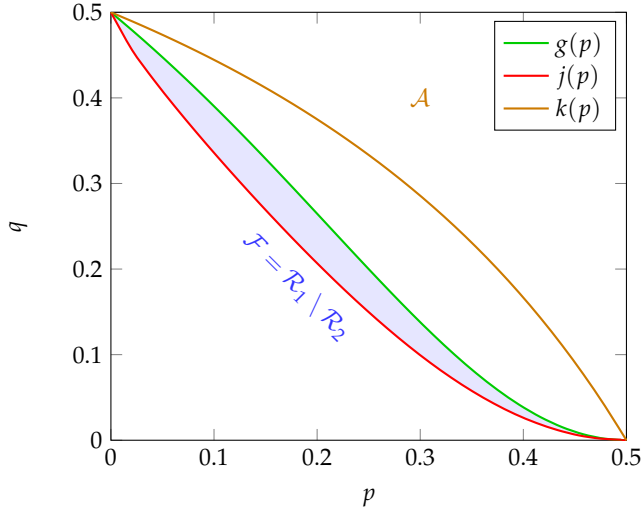

We now show that in a subregion of $\cR_1$ the optimizing state for $I_c(\Npq)$ is the completely mixed state $\pi\coloneqq \frac{1}{2}\one$, corresponding to $z=0$.
Observe that
\begin{align}
i_c'(z) = \frac{1}{2 k(p,z)}\left( (1-2q)k(p,z) \log\frac{1-z}{1+z} - 4p(1-p)(1-q)z \log\frac{1-k(p,z)}{1+k(p,z)} \right),
\end{align}
where $k(p,z)\coloneqq \sqrt{1-4p(1-p)(1-z^2)}$ satisfying $k(p,z)>0$ for all $z\in[-1,1]$ and $p\in[0,1/2)$.
It is easy to see that $i_c'(0)=0$.
Evaluating the second derivative of $i_c(z)$ at $z=0$ gives
\begin{align}
i_c''(0) = -(\ln 2)^{-1}(1-2q) + \frac{2p(1-p)}{1-2p}(1-q)\log\frac{1-p}{p},
\end{align}
which satisfies $i_c''(0)<0$ if 
\begin{align}
q < j(p)\coloneqq \frac{1-2p-2p(1-p)\ln\frac{1-p}{p}}{2-4p-2p(1-p)\ln\frac{1-p}{p}}.\label{eq:j-app}
\end{align}
In Fig.~\ref{fig:fish} we plot the regions $\cA$ and $\cR_1$ defined in \eqref{eq:A-app} and \eqref{eq:R1-app}, respectively, together with the regions 
\begin{align}
\cR_2 &\coloneqq \left\lbrace (p,q)\colon p\in[0,1/2],\, 0 \leq q < j(p)\right\rbrace\label{eq:R2-app}\\
\cF &\coloneqq \cR_1\setminus\cR_2 = \left\lbrace (p,q)\colon p\in[0,1/2],\, j(p)\leq q <g(p) \right\rbrace.\label{eq:F}
\end{align}

\section{Superadditivity of coherent information}\label{app:superadditivity}
In the following, we show that in a certain subregion of the $(p,q)$-plane a simple weighted $n$-repetition code 
\begin{align}
\rho_n \coloneqq \lambda |0\rangle\langle 0|^{\ox n} + (1-\lambda) |1\rangle\langle 1|^{\ox n},
\label{eq:repetition-code-app}
\end{align}
achieves superadditivity of the coherent information, $I_c(\rho_n,\Npq^{\ox n}) > n I_c(\Npq)$. 

First, we note that the coherent information of $\Npq$ can be rewritten as
\begin{align}
I_c(\rho_n,\Npq^{\ox n}) = S\left(\Npq^{\ox n}(\rho_n)\right) - S\left((\id_2\ox\Npq^{\ox n})(\phi_n)\right),\label{eq:coh-info-purification-app}
\end{align}
where $|\phi_n\rangle \coloneqq \sqrt{\lambda} |0\rangle^{\ox n+1} + \sqrt{1-\lambda} |1\rangle^{\ox n+1}$ is a purification of $\rho_n$.
In the following, we denote by $\cZ_p(\rho) = (1-p)\rho + pZ\rho Z$ the dephasing channel, and by $\cE_q(\rho)=(1-q)\rho + q\tr(\rho)|e\rangle\langle e|$ the erasure channel.
Setting $q_0 = 1-q$ and $q_1 = q$, as well as $\cC_0=\cZ_p$ and $\cC_1=\cE_1$, we have
\begin{align}
\Npq^{\ox n} &= ((1-q)\cZ_p + q \cE_1)^{\ox n} \\
&= \sum_{s^n\in \lbrace 0,\,1\rbrace^n} q_{s^n} \cC_{s^n},
\end{align}
where $s^n\in \lbrace 0,\,1\rbrace^n$ denotes a binary string of length $n$, and we set $q_{s^n}\coloneqq \prod_{j=1}^n q_{s_j}$ and $\cC_{s^n}\coloneqq \bigotimes_{j=1}^n \cC_{s_j}$.
Note that $\cC_{s^n}(\sigma_n) \perp \cC_{t^n}(\sigma_n)$ for $s^n\neq t^n$ and any $\sigma_n$ due to the different erasure patterns originating from the action of $\cE_1$ on different tensor factors.
Hence, by \eqref{eq:entropy-cq-factorization} we have
\begin{align}
I_c(\rho_n,\Npq^{\ox n}) = \sum_{s^n\in\lbrace 0,\,1\rbrace^n} q_{s^n} \left[ S(\cC_{s^n}(\rho_n)) - S((\id_2\ox \cC_{s^n})(\phi_n)) \right].
\label{eq:coh-info-split-up}
\end{align}
Let $s^n\notin \lbrace (0,\dots,0), (1,\dots,1)\rbrace$.
Then, after the erasure on the systems $\lbrace j\colon s_j=1\rbrace$,
\begin{align}
\cC_{s^n}(\rho_n) &= \lambda \bigotimes\nolimits_{j\colon s_j=0}\cZ_p(|0\rangle\langle 0|) + (1-\lambda)\bigotimes\nolimits_{j\colon s_j=0}\cZ_p(|1\rangle\langle 1|)\\
&= \lambda \bigotimes\nolimits_{j\colon s_j=0} |0\rangle\langle 0| + (1-\lambda) \bigotimes\nolimits_{j\colon s_j=0} |1\rangle\langle 1|,
\intertext{and similarly,}
\cC_{s^n}(\phi_n)&= \lambda |0\rangle\langle 0|\ox \bigotimes\nolimits_{j\colon s_j=0}\cZ_p(|0\rangle\langle 0|) + (1-\lambda)|1\rangle\langle 1|\ox \bigotimes\nolimits_{j\colon s_j=0}\cZ_p(|1\rangle\langle 1|)\\
&= \lambda |0\rangle\langle 0|\ox \bigotimes\nolimits_{j\colon s_j=0} |0\rangle\langle 0| + (1-\lambda) |1\rangle\langle 1| \ox \bigotimes\nolimits_{j\colon s_j=0} |1\rangle\langle 1|.
\end{align}
Hence, $S(\cC_{s^n}(\rho_n)) = S(\cC_{s^n}(\phi_n))$ for all strings $s^n\notin \lbrace (0,\dots,0), (1,\dots,1)\rbrace$, and by \eqref{eq:coh-info-split-up} the only terms contributing to $I_c(\rho_n,\Npq^{\ox n})$ are those with $\cC_{(0,\dots,0)} = \cZ_p^{\ox n}$ and $\cC_{(1,\dots,1)} = \cE_1^{\ox n}$.
For the latter, note that $S(\cE_1^{\ox n} (\rho_n)) = 0$ and $S(\cE_1^{\ox n}(\phi_n)) = S(\lambda |0\rangle\langle 0| + (1-\lambda)|1\rangle\langle 1| ) = h(\lambda)$.

Up to now, we have shown that
\begin{align}
I_c(\rho_n,\Npq^{\ox n}) = (1-q)^n \left(S(\cZ_p^{\ox n}(\rho_n)) - S((\id_2\ox\cZ_p^{\ox n})(\phi_n)) \right) - q^n h(\lambda).
\end{align}
We have $\cZ_p^{\ox n}(\rho_n) = \rho_n$, and hence $S(\cZ_p^{\ox n}(\rho_n)) = h(\lambda)$.
It thus remains to compute $S((\id_2\ox\cZ_p^{\ox n})(\phi_n))$.
To this end, observe that (omitting identity maps and identity operators)
\begin{align}
\cZ_p^{\ox n} (\phi_n) = \sum_{s^n\in\lbrace 0,\,1\rbrace^n} p_{s^n} Z_{s^n} |\phi_n\rangle\langle\phi_n| Z_{s^n},
\end{align}
where we again set $p_{s^n}\coloneqq \prod_{j=1}^n p_{s_j}$ and $Z_{s^n}\coloneqq \bigotimes_{j=1}^n Z^{s_j}$ (with $Z^0=\one$).
We have $Z_{s^n} |\phi_n\rangle = |\phi_n\rangle$ if $|s^n|$ is even, and $Z_{s^n} |\phi_n\rangle = |\tphi_n\rangle \coloneqq \lambda |0\rangle^{\ox n+1} - (1-\lambda) |1\rangle^{\ox n+1}$ if $|s^n|$ is odd. 
Then,
\begin{align}
\cZ_p^{\ox n}(\phi_n) &= \sum_{s^n\colon |s^n| \text{ even}} p_{s^n} |\phi_n\rangle\langle\phi_n| + \sum_{s^n\colon |s^n| \text{ odd}} p_{s^n} |\tphi_n\rangle\langle\tphi_n|\\
&= \frac{1}{2}(1+(1-2p)^n) |\phi_n\rangle\langle\phi_n| + \frac{1}{2}(1-(1-2p)^n) |\tphi_n\rangle\langle\tphi_n|\\
&= \begin{bmatrix} \lambda & (1-2p)^n\sqrt{\lambda(1-\lambda)} \\ (1-2p)^n\sqrt{\lambda(1-\lambda)} & 1-\lambda \end{bmatrix}_{\lbrace |0\rangle^{\ox n+1},\,|1\rangle^{\ox n+1}\rbrace}\\
&\eqqcolon M_{\lambda,\,p},
\end{align}
where we used $\sum_{\,s^n\colon |s^n| \text{ even}} p_{s^n} =  \frac{1}{2}(1+(1-2p)^n)$ and $\sum_{\,s^n\colon |s^n| \text{ odd}} p_{s^n} = \frac{1}{2}(1-(1-2p)^n)$ in the second equality.
The matrix $M_{\lambda,\,p}$ is written out with respect to the basis ${\lbrace |0\rangle^{\ox n+1},\,|1\rangle^{\ox n+1}\rbrace}$.
Its von Neumann entropy evaluates to
\begin{align}
S(M_{\lambda,\,p}) = 1-u(\lambda,p,n)\artanh u(\lambda,p,n) - \frac{1}{2}\log\left(1-u(\lambda,p,n)^2\right),
\end{align}
where $u(\lambda,p,n)\coloneqq \sqrt{1- 4\lambda(1-\lambda)(1-(1-2p)^{2n})}$ and $\artanh(x) \coloneqq \frac{1}{2}\log\frac{1+x}{1-x}$.

\begin{figure}[ht]
	\centering
	\includegraphics[width=\textwidth]{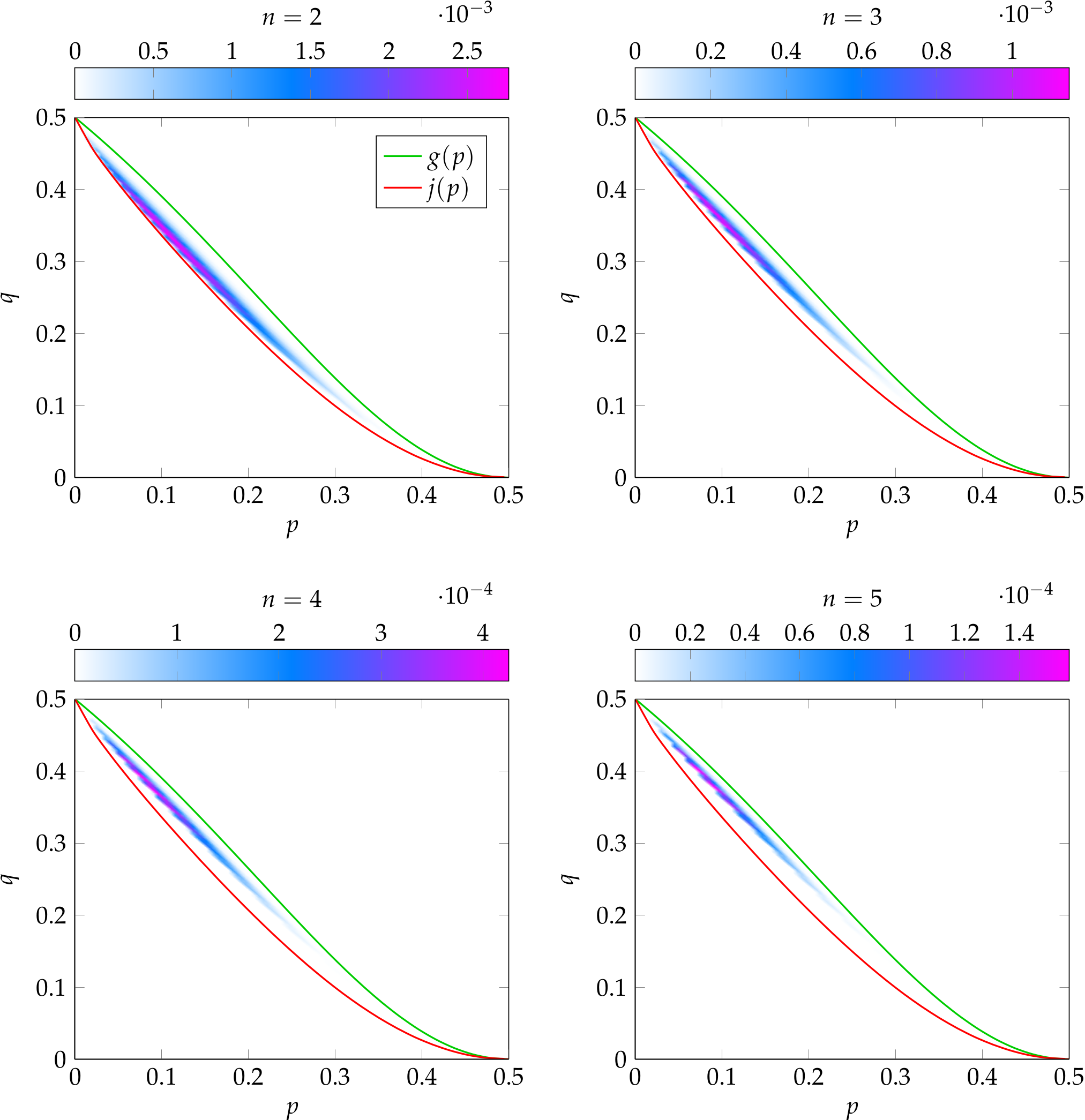}
	\caption{Plot of the non-negative part of $\max_\lambda I_c(\rho_n,\Npq^{\ox n})/n - I_c(\Npq)$ for $n\in\lbrace 2,3,4,5\rbrace$ (top-left to bottom-right), where $\rho_n$ is the $n$-repetition code defined in \eqref{eq:repetition-code-app}. 
		Note the different scaling of the colorbars in both plots, indicating that the magnitude of superadditivity of the coherent information decreases with $n$.
		The functions $g(p)$ (green) and $j(p)$ (red) are defined in \eqref{eq:g-app} and \eqref{eq:j-app}, respectively.
	}
	\label{fig:repcodes-heatmap}
\end{figure}

In summary, we have
\begin{align}
I_c(\rho_n,\Npq^{\ox n}) = ((1-q)^n-q^n)h(\lambda) - (1-q)^n\left(1-u(\lambda,p,n)\artanh u(\lambda,p,n) - \frac{1}{2}\log\left(1-u(\lambda,p,n)^2\right)\right).
\label{eq:coh-info-repetition-code-app}
\end{align}
In Fig.~\ref{fig:repcodes-heatmap}, we plot heat maps of the quantity $\max_\lambda I_c(\rho_n,\Npq^{\ox n})/n - I_c(\Npq)$ for $n\in\lbrace 2,3,4,5\rbrace$, showing examples of superadditivity of coherent information of the dephrasure channel.

The fact that the single-letter coherent information $I_c(\Npq)$ is maximized by states diagonal in the $Z$-basis suggests to consider more general $Z$-diagonal codes whose Schmidt rank is larger than 2, generalizing the weighted repetition code defined in \eqref{eq:repetition-code-app}.
To this end, for $n\geq 2$ we fix the Schmidt basis $\left\lbrace |s^n\rangle \ox |s^n\rangle\right\rbrace_{s^n\in\lbrace 0,1\rbrace^n}$, where $s^n$ ranges over all binary strings of length $n$, and optimize the Schmidt coefficients $\lbrace\lambda_{s^n}\rbrace_{s^n\in\lbrace 0,1\rbrace^n}$ in
\begin{equation}
|\theta_n\rangle = \sum_{s^n\in\lbrace 0,1\rbrace^n} \lambda_{s^n} |s^n\rangle \ox |s^n\rangle. \label{eq:dephased-code}
\end{equation}
Here, the $n$-fold dephrasure channel $\Npq^{\ox n}$ acts on the $n$ qubits in the right-hand tensor factor.
A straightforward optimization over codes of the form \eqref{eq:dephased-code} now yields codes that outperform both the optimal single-letter code as well as the weighted repetition code \eqref{eq:repetition-code-app}.
In Fig.~\ref{fig:repcodes} we plot the optimized $Z$-diagonal code $|\theta_4\rangle$ on 4 input qubits, which provides a good trade-off between rate and computational cost of optimization.

However, we were also able to find \emph{non-diagonal} codes that outperform the dephased codes $|\theta_n\rangle$ for certain values of $(p,q)$.
These codes were found using a full parametrization of the code state and a global optimization technique called \emph{particle swarm optimization} (see the last section below for a high-level explanation of this technique).
In particular, we found the following code $\chi_3$ for $3$ channel uses:
\begin{eqnarray}
|\chi_3\rangle \coloneqq |0000\rangle \ox |\psi_1\rangle + |1111\rangle \ox |\psi_1\rangle + |0101\rangle \ox |\psi_2\rangle + |1010\rangle \ox X|\psi_2\rangle, \label{eq:chi3}
\end{eqnarray}
where $|\psi_i\rangle \coloneqq c_i |0\rangle + d_i |1\rangle$, and $X=|0\rangle\langle 1| + |1\rangle\langle 0|$ denotes the Pauli $X$-operator.
All coefficients above are chosen such that the code state $\chi_3$ is normalized.
Feeding the $3$ right-most qubits of $|\chi_3\rangle$ into the $3$-fold dephrasure channel $\cN_{p,\,q}^{\ox 3}$, we optimized the coefficients $c_i$, and $d_i$ in order to maximize the coherent information $I_c(\chi_3,\Npq^{\ox 3})$.
The resulting rates along the $(p,3p)$-diagonal for the interval $p\in[0.107,0.118]$ are shown in Fig.~\ref{fig:repcodes}.
We also note that other interesting non-diagonal codes can be obtained using a neural network state ansatz \cite{BL18}.

\section{Threshold of the weighted repetition codes}\label{app:threshold}
In this appendix, we show that the threshold of the coherent information $I_c(\rho_n,\Npq^{\ox n})$ of the weighted repetition code \eqref{eq:repetition-code} coincides with the fuction $g(p)$ defined in \eqref{eq:g} for all $n\in\mathbb{N}$.
To this end, we first rewrite the formula \eqref{eq:coh-info-repetition-code} for $I_c(\rho_n,\Npq^{\ox n})$ in a slightly different way:
\begin{align}
I_c(\rho_n,\Npq^{\ox n}) = ((1-q)^n-q^n) h(\lambda) - (1-q)^n \left( 1 - \frac{1+u}{2}\log(1+u) - \frac{1-u}{2}\log(1-u)\right),
\end{align}
where $u=u(p,\lambda,n) = \sqrt{1-4\lambda(1-\lambda)(1-(1-2p)^{2n})}$.

 Recall that $\rho_n = \lambda |0\rangle\langle 0|^{\ox n} + (1-\lambda) |1\rangle\langle 1|^{\ox n}$.
 To determine all $(p,q)$ such that $I_c(\rho_n,\Npq^{\ox n})>0$, let us assume $\lambda>0$ such that $h(\lambda)>0$.
Then, $I_c(\phi_n,\Npq^{\ox n})>0$ if and only if 
\begin{align}
\frac{(1-q)^n-q^n}{(1-q)^n} = 1 - \left(\frac{q}{1-q}\right)^n >  \frac{1 - \frac{1+u}{2}\log(1+u) - \frac{1-u}{2}\log(1-u)}{h(\lambda)} \eqqcolon f(p,\lambda,n).\label{eq:threshold}
\end{align}
We show below that, for all $n\in\mathbb{N}$, we have $f(p,\lambda,n) \searrow 1-(1-2p)^{2n}$ as $\lambda\searrow 0$.
Hence, for all $n\in\mathbb{N}$ there is a $\lambda>0$ such that \eqref{eq:threshold} holds provided that
\begin{align}
1 - \left(\frac{q}{1-q}\right)^n > 1-(1-2p)^{2n},
\end{align}
which is equivalent to $q < \frac{(1-2p)^2}{1+(1-2p)^2}  =g(p)$.

It remains to be shown that $f(p,\lambda,n) \searrow 1-(1-2p)^{2n}$ for all $n\in\mathbb{N}$.
We assume that $\lambda > 0$ is small, and use the following approximations, abbreviating $c_{p,n}\coloneqq 1-(1-2p)^{2n}$:
\begin{align}
u(p,\lambda,n) &= 1 - 2\lambda(1-\lambda)c_{p,n} + O(\lambda^2)\\
-(1-\lambda)\log(1-\lambda) &= (\ln 2)^{-1} \lambda + O(\lambda^2)\\
-(1-\lambda(1-\lambda)c_{p,n})\log (1-\lambda(1-\lambda)c_{p,n}) &= (\ln 2)^{-1} c_{p,n} \lambda + O(\lambda^2).
\end{align}
Consider now that
\begin{align}
&1-\frac{1+u}{2}\log(1+u) - \frac{1-u}{2}\log(1-u)\\
&\qquad\qquad {} = 1-(1-\lambda(1-\lambda)c_{p,n})\log[2(1-\lambda(1-\lambda)c_{p,n})] - \lambda(1-\lambda)c_{p,n}\log[2\lambda(1-\lambda)c_{p,n}]\\
&\qquad\qquad {} = -(1-\lambda(1-\lambda)c_{p,n})\log[1-\lambda(1-\lambda)c_{p,n}] - \lambda(1-\lambda)c_{p,n}\log[\lambda(1-\lambda)c_{p,n}]\\
&\qquad\qquad {} = (\ln 2)^{-1} c_{p,n}\lambda - c_{p,n} \lambda \log\lambda - \lambda c_{p,n} \log c_{p,n}\\
&\qquad\qquad {} = \lambda c_{p,n} ((\ln 2)^{-1} - \log\lambda - \log c_{p,n}),
\end{align}
where we used the above approximations and neglected occurrences of $\lambda^2$.
We also have $h(\lambda) = -\lambda\log\lambda + (\ln 2)^{-1}\lambda + O(\lambda^2)$.
Hence,
\begin{align}
f(p,\lambda,n) &= \frac{1 - \frac{1+u}{2}\log(1+u) - \frac{1-u}{2}\log(1-u)}{h(\lambda)}\\
&= c_{p,n}\left(1 - \frac{\log c_{p,n}}{(\ln 2)^{-1} - \log\lambda}\right).
\end{align}
Evidently, $\lim_{\lambda\to 0} f(p,\lambda,n) = c_{p,n}$, and this convergence is monotonic, since $c_{p,n}=1-(1-2p)^{2n} \in(0,1]$ for all $n\in\mathbb{N}$ (note that $p\in[0,1/2]$).
Hence, $\log c_{p,n} < 0$, and $\lambda\mapsto 1 - \frac{\log c_{p,n}}{(\ln 2)^{-1} - \log\lambda}$ is monotonically increasing in $\lambda$.

\section{Separation of private information and coherent information}\label{app:separation}
To show a separation between private information and coherent information, we searched for private codes $\kE=\lbrace p_x, \rho_x\rbrace$ by first considering a pure state $|\psi\rangle_{SRA}$, where $R$ is a purifying system for $A$ with $|R|=|A|$, and $|S|=|A|^2$ is an auxiliary system. 
To obtain a private code $\lbrace p_x, \rho_x\rbrace$, we first measure $S$ with respect to the computational basis to obtain $|A|^2$ pure states $|\phi_x\rangle_{RA}$, where $x=1, \dots,|A|^2$. 
We then set $p_x = \langle \phi_x|\phi_x\rangle$ and $\rho_x = \frac{1}{p_x}\tr_R \phi_x$.
This simple procedure leads to private codes that can be symmetrized to one of the form as in \eqref{eq:opt-private-ensemble}.
Extensive numerical search furthermore suggests the optimality of these private codes.

\section{Coherent information of the complementary channel}\label{app:coh-info-complementary}

In this appendix, we show that the complementary channel $\Npq^c$ has positive coherent information for all $p,q\in[0,1/2]$, and hence, $0 < Q(\Npq^c) \leq P(\Npq^c)$ for all $p,q\in[0,1/2]$.
We prove this by constructing a particular input state $\rho$ satisfying $I_c(\rho,\Npq^c) >0$.

Since the complementary channel \eqref{eq:complementary-channel-app} is flagged, it follows that, using \eqref{eq:entropy-cq-factorization},
\begin{align}
I_c(\rho,\Npq^c) = q S(\rho) + (1-q) \left[S(\cZ_p^c(\rho)) - S(\cZ_p(\rho))\right]\quad\text{for all $\rho$.}\label{eq:comp-coh-info}
\end{align}
We now choose $\rho = \frac{1}{2}( I + m X)$ with $m\in[0,1]$, for which $S(\rho) = h(\eps)$ with $\eps=\frac{1-m}{2}$.
Simple calculations further show that
\begin{align}
\cZ_p(\rho) &= \frac{1}{2}(I + (1-2p)mX),\\
S(\cZ_p(\rho)) &= h(\eps + p - 2\eps p).
\end{align}
Moreover, $\langle 0|\rho|0\rangle=\langle 1|\rho|1\rangle=\frac{1}{2}$, and hence 
\begin{align}
\cZ_p^c(\rho) &= (1-p)|0\rangle\langle 0| + p|1\rangle\langle 1|,\\
S(\cZ_p^c(\rho)) &= h(p).
\end{align}

Substituting the above in \eqref{eq:comp-coh-info} gives
\begin{align}
I_c(\rho,\Npq^c) &= q h(\eps) + (1-q) \left[ h(p) - h(p+\eps - 2\eps p)\right]\\
&= q h(\eps) - (1-q) \eps (1-2p) \frac{h(p+\eps - 2\eps p) - h(p)}{\eps(1-2p)}\\
&= q h(\eps) - (1-q) \eps (1-2p)\, h'(x)\bigr|_{x=\bar{p}}\\
&\eqqcolon q h(\eps) - (1-q) \eps (1-2p) C_{\bar{p}},\label{eq:cci-C}
\end{align}
where in the third equality we applied the mean value theorem for the difference quotient with a suitable $\bar{p}\in (p,p+\eps-2\eps p)$, and in the fourth equality we defined $C_{\bar{p}}\coloneqq h'(x)\bigr|_{x=\bar{p}} = \log\frac{1-\bar{p}}{\bar{p}}$.
From \eqref{eq:cci-C}, we have $I_c(\rho,\Npq^c)>0$ if and only if 
\begin{align}
\frac{h(\eps)}{\eps} > \frac{1-q}{q} (1-2p) C_{\bar{p}}. \label{eq:condition}
\end{align}
Note that $\frac{h(\eps)}{\eps}\to\infty$ in the limit $\eps\to 0$.
Therefore, for all $p,q$ there exists an $\eps>0$ such that \eqref{eq:condition} holds true.

To obtain an explicit expression for $\eps$, we use the bound $\frac{h(\eps)}{\eps} > -\log \eps$ to see that any $\eps>0$ satisfying $-\log \eps > \frac{1-q}{q} (1-2p) C_{\bar{p}}$, or
$0 < \eps < \exp\left[-\frac{1-q}{q} (1-2p) C_{\bar{p}}\right],$
is sufficient to assert \eqref{eq:condition}.
Note that $C_{\bar{p}} \nearrow h'(x)\bigr|_{x=p} = \log\frac{1-p}{p}$ in the limit $\eps\to 0$ due to concavity of $h(\cdot)$.
Hence, any $\rho = \frac{1}{2}( I + m X)$ with $\eps=\frac{1-m}{2}$ satisfying
\begin{align}
0 < \eps < \exp\left[-\frac{1-q}{q} (1-2p) \log\frac{1-p}{p}\right]
\end{align}
yields $I_c(\rho,\Npq^c) > 0$.

\section{Particle swarm optimization}\label{app:particle-swarm}
In this section we provide a high-level description of \emph{particle swarm optimization} \cite{KE95}, a global optimization technique that we used to optimize the coherent information of the dephrasure channel.

For the sake of simplicity, assume that we want to find the global minimum of a function $f\colon \mathbb{R}^d \to \mathbb{R}$.
To this end, we first fix three parameters $c_I,c_\text{self},c_\text{soc}\geq 0$.
We then send out $n$ \emph{agents} or \emph{particles}, each initialized at a random location $\mathbf{x}_i\in\mathbb{R}^d$ where $i=1,\dots,n$, and evaluate the target function $f(\mathbf{x}_i)$.
In subsequent iterations, the ``velocity'' $\mathbf{v}_i$ of the $i$-th particle is updated according to the rule
\begin{align}
\mathbf{v}_i \leftarrow c_I \mathbf{v}_i + c_\text{self} u_\text{self} (\mathbf{x}_i - \mathbf{p}_i) + c_\text{soc} u_\text{soc} (\mathbf{x}_i - \mathbf{g}),\label{eq:update-rule}
\end{align}
where $u_\text{self},u_\text{soc}\in[0,1]$ are drawn uniformly at random, the vector $\mathbf{p}_i$ holds the location of the minimal value of $f$ that the $i$-th agent has seen, and the vector $\mathbf{g}$ holds the location of the minimal value of $f$ \emph{any} agent in the group has seen.
In other words, in each iteration the velocity of an agent is updated according to the current position (governed by the ``inertia'' term $c_I$), the position of its personal best value of $f$ (governed by the ``self-interaction'' term $c_\text{self}$), and the position of the group best value of $f$ (governed by the ``social-interaction'' term $c_\text{soc}$).
In the beginning, the velocity of each particle is initialized uniformly at random.

Particle swarm optimization does not make use of the gradient of $f$, and is thus suitable for unstructured optimization problems where differentiability of the target function $f$ is not known.
Moreover, the update rule \eqref{eq:update-rule} allows individual agents to escape local minima, thus ensuring a more thorough exploration of the landscape (visualizing multiple iteration steps in a successful optimization run, the group of agents slowly gravitates towards the global minimum, much like a swarm of insects gravitating towards a food source).

The coherent information of a quantum channel has many such local extrema, as any pure input state yields zero coherent information.
This suggests the use of gradient-free optimization algorithms such as particle swarm optimization to find improved local (or even global) extrema of the coherent information.
Using particle swarm optimization, we were able to find improved quantum codes for the dephrasure channel (such as the code $\chi_3$ defined in \eqref{eq:chi3}) that yield higher rates than the simple repetition code $\rho_n$ (albeit with a lower threshold).
Presumably, particle swarm optimization might also be successfully applied to other optimization problems in quantum information theory for which the existence of a large number of local extrema renders gradient-based methods inefficient.
\end{document}